\documentclass[preprint]{aastex}
\usepackage{graphicx}

\begin{document}
\title{A Giant Flare from a Soft Gamma Repeater in the Andromeda
Galaxy, M31}

\author{E.~P.~Mazets\altaffilmark{1}, R.~L.~Aptekar\altaffilmark{1},
T.~L.~Cline\altaffilmark{2},  D.~D.~Frederiks\altaffilmark{1},
J.~O.~Goldsten\altaffilmark{3}, S.~V.~Golenetskii\altaffilmark{1},
K.~Hurley\altaffilmark{4}, A.~von~Kienlin\altaffilmark{5}, and
V.~D.~Pal'shin\altaffilmark{1}}

\altaffiltext{1}{Ioffe Physico-Technical Institute of the Russian
Academy of Sciences, St. Petersburg, 194021, Russia}
\altaffiltext{2}{Goddard Space Flight Center, NASA, Greenbelt, MD
20771, USA}
\altaffiltext{3}{The Johns Hopkins University Applied Physics
Laboratory, MD 20723, USA}
\altaffiltext{4}{Space Sciences Laboratory, University of California
at Berkeley, Berkeley, CA 94720-7450, USA}
\altaffiltext{5}{Max-Plank-Institut f\"{u}r extraterrestrische
Physik, D-85741 Garching, Germany}

\keywords{gamma-ray bursts; soft gamma-ray repeaters}

\begin{abstract}
The light curve, energy spectra, energetics, and IPN localization of
an exceedingly intense short duration hard spectrum burst,
GRB~070201,  obtained from Konus-Wind, INTEGRAL (SPI-ACS), and
MESSENGER data are presented. The total fluence of the burst and the
peak flux are $S = 2.00_{-0.26}^{+0.10} \times
10^{-5}$~erg~cm$^{-2}$ and $F_{\mathrm{max}} = 1.61_{-0.50}^{+0.29}
\times 10^{-3}$~erg~cm$^{-2}$~s$^{-1}$. The IPN error box has an
area of 446 square arcminutes and covers the peripheral part of the
M31 galaxy. Assuming that the source of the burst is indeed in M31
at a distance of 0.78 Mpc, the measured values of the fluence $S$
and maximum flux $F_{\mathrm{max}}$ correspond to a total energy of
$Q = 1.5 \times 10^{45}$~erg, and a maximum luminosity $L = 1.2
\times 10^{47}$~erg~s$^{-1}$. These data are in good agreement with
the corresponding characteristics of the previously observed giant
flares from other soft gamma repeaters. The evidence for the
identification of this event as a giant flare from a soft gamma
repeater in the M31 galaxy is presented.
\end{abstract}

\section{INTRODUCTION}
Soft gamma repeaters (SGRs) are a special rare class of strongly
magnetized neutron stars exhibiting two types of gamma-ray burst
emission. Occasionally, SGRs enter an active stage and emit repeated
short bursts with spectra which in an energy range above 10--15~keV
can be approximated by a soft thermal bremsstrahlung-like function
with $kT \approx 20$--30~keV. The bursting activity  may last from
several days to a year or more,  followed by a long quiescent period
lasting up to several years.

Much more rarely, perhaps once every 30--40 years, an SGR may emit a
giant flare (GF) with enormous intensity. The energy released in
such a flare in gamma-rays is comparable to the total energy emitted
by the Sun over 10$^4$--10$^5$ years or more. GFs display short
($\sim 0.2$--0.5~s) initial pulses of hard gamma rays with a steep
leading edge and a more extended trailing edge which  as a rule
evolves into a soft long decaying tail pulsating with the neutron
star rotation period. A GF is frequently preceded by pronounced
increase in bursting  activity \citep{Frederiks2007a}.

The first two SGRs were discovered and localized in March, 1979 by
the Konus experiment on Venera 11 and 12. A giant flare from
SGR~0526-66 was detected on March 5, 1979 by numerous spacecraft
 \citep{Mazets1979a, Evans1980, Cline1980}. One
of the first results of the interplanetary network (IPN) was the
localization of the first GF to an error box inside the N49
supernova remnant in the Large Magellanic Cloud (LMC)
\citep{Cline1980}. Another important result was the observation and
localization by the Konus experiment of $\sim$17 weak soft recurrent
bursts emitted by SGR~0526-66 in the years following  the GF
\citep{Mazets1979a, Golenetskii1984}. Moreover, in 1979 March, Konus
detected and localized  three similar recurrent bursts from another
source, SGR~1900+14 \citep{Mazets1979b}.  The third SGR, 1806-20,
was discovered in 1983 \citep{Laros1987, Atteia1987,
Kouveliotou1987}, and the fourth, SGR~1627-41, in 1998
\citep{Hurley1999b, Woods1999}. It has become clear that SGRs are
very rare astrophysical objects. The giant flare on 1979 March 5
remained a unique event for 20 years, until new giant flares were
detected from SGR~1627-41 on 1998 June 18 \citep{Mazets1999a} and
from SGR~1900+14 on 1998 August 27 \citep{Hurley1999a, Feroci1999,
Mazets1999b}. Finally, a giant flare from SGR~1806-20 on December
27, 2004  was also observed \citep{Hurley2005, Frederiks2007a}. The
extremely high intensity of the initial pulses of the giant flares
suggested that these events could be detected in galaxies at
distances up to 10--30~Mpc.

The short hard GRB~051103 was localized by the interplanetary
network to the nearby M81 group of interacting galaxies which lies
at a distance of 3.6~Mpc \citep{Golenetskii2005}.  The possibility
of identifying this burst as a GF from an SGR in the M81 group, as
opposed to a short GRB in a background galaxy, has been discussed
\citep{Frederiks2007b, Hurley2007a}.

On 2007 February 1 Konus-Wind  recorded a short duration hard
spectrum gamma-ray burst which had a higher intensity than  any
previously observed event. The burst was localized by the IPN
\citep{Golenetskii2007}. The final localization (a 446 arcmin$^2$
error box) covers the outer arms of the M31 galaxy. The analysis
presented below of the time history, spectral characteristics, and
energetics argues strongly for the interpretation that this event is
actually a GF from a soft gamma repeater in the Andromeda galaxy. In
a companion paper, \citet{Abbott2007} discuss the LIGO data taken at
the time of this event.

\section{OBSERVATIONS AND LOCALIZATION}
\begin{figure}
\centering
\includegraphics[]{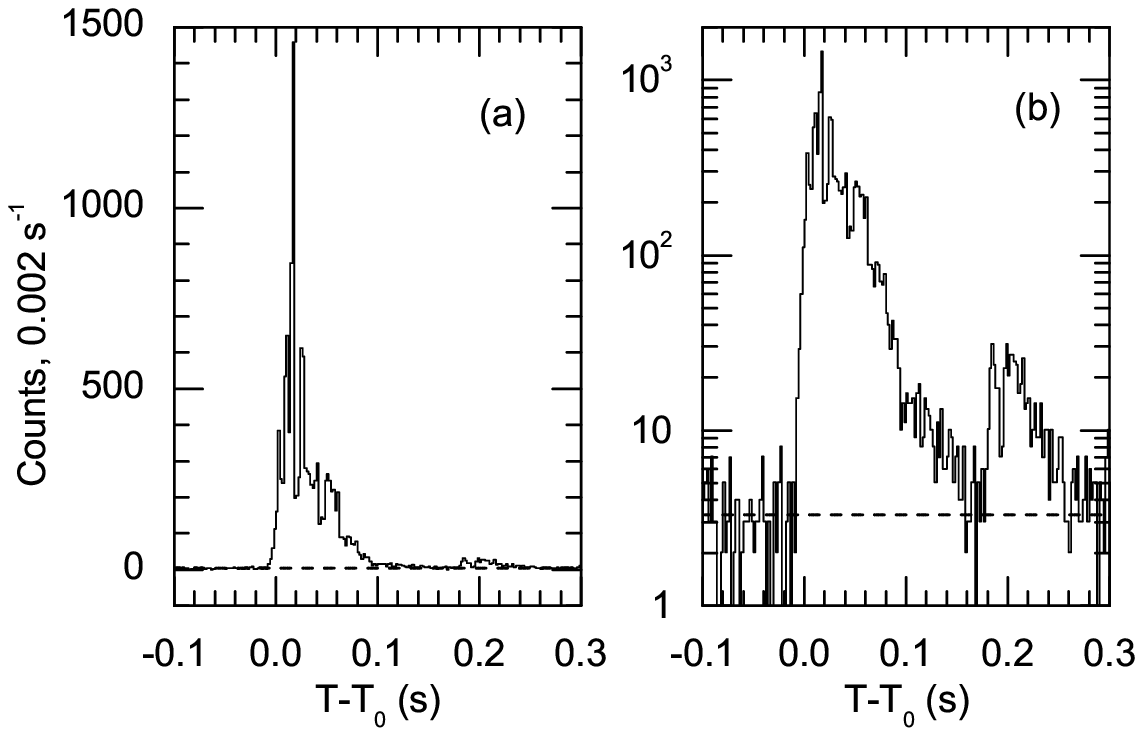}
\caption{17--1130 keV Konus-Wind light curve of GRB~070201 (dead
time corrected) at the highest available  time resolution (2 ms)
presented on a linear count rate scale (a), and also on a
logarithmic scale (b) to enhance the  low-intensity portions.
\label{lcSum}}
\end{figure}

\begin{figure}
\centering
\includegraphics[]{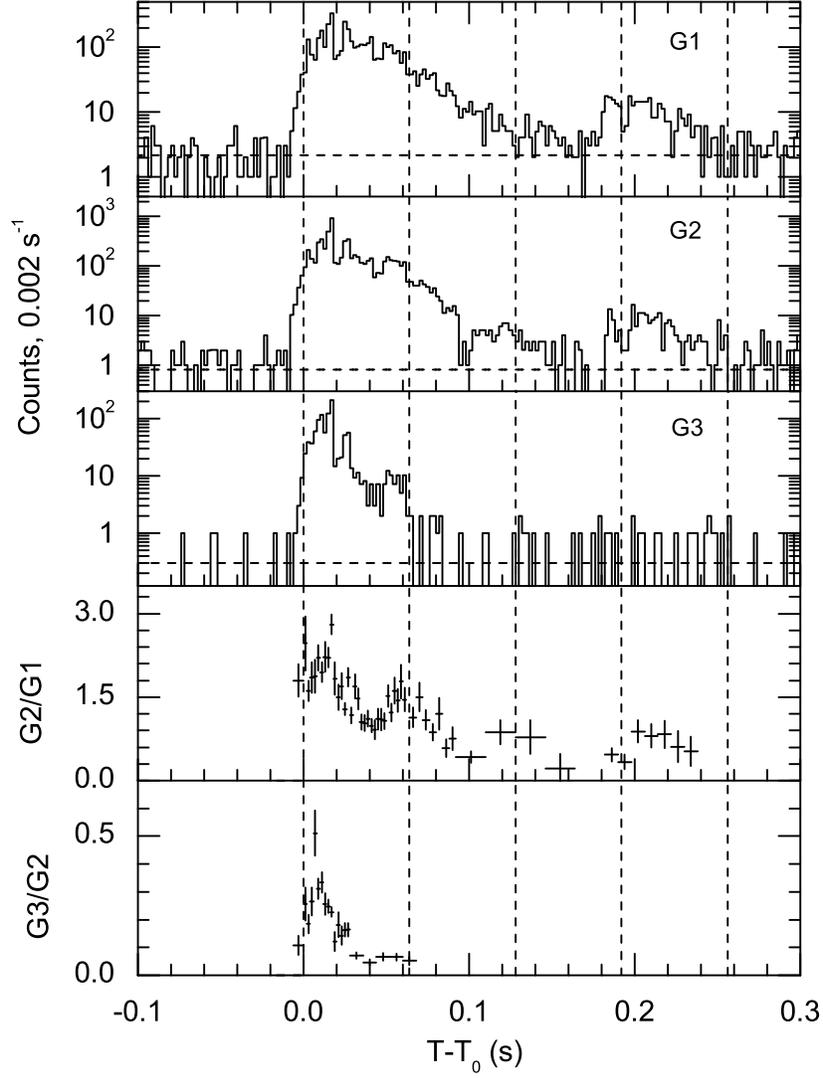}
\caption{Konus-Wind light curve in three energy bands: G1 (17--70
keV), G2 (70--300 keV), and G3 (300--1130 keV), and the hardness
ratios G2/G1 and G3/G2. The vertical dashed lines indicate the four
successive 64-ms intervals where the energy spectra were measured.
\label{lc3bands}}
\end{figure}

The light curve of GRB~070201 (Konus trigger time T$_0$=55390.780~s
UT, corresponding to an Earth-crossing time 55390.261~s) recorded by
the Konus-Wind S2 detector in the energy range 17--1130~keV is shown
in Fig.~\ref{lcSum}. It displays a narrow pulse with a rather steep
leading edge ($\sim$20~ms) followed by more prolonged decay to
T-T$_0 \approx 180$~ms. The maximum count rate occurs in a $\leq
2$~ms long interval. The burst profile is not smooth: in most cases,
the variations evident in the count rate at the top of the pulse and
in its decay phase are statistically significant. A weak secondary
flash is observed at T-T$_0 \approx 180$--280~ms. Comparison of the
light curves in three energy bands G1 (17--70~keV), G2
(70--300~keV), and G3 (300--1130~keV) (Fig.~\ref{lc3bands}) and of
the hardness ratios G2/G1 and G3/G2 reveals a strong, rapid spectral
evolution of the radiation. The most pronounced changes are observed
for the high-energy part of the spectrum (energies $E_\gamma >
300$~keV). The duration of the high-energy radiation does not exceed
80 ms. The second, weak rise in intensity at T-T$_0 \approx 180$~ms
is present only for the soft part of the spectrum at energies
$E_\gamma < 300$~keV. Four multichannel spectra were measured in the
course of the burst, each with an accumulation time of 64~ms. They
cover the energy range from 17~keV to 14~MeV, but no statistically
significant emission is seen above 2~MeV.  The boundaries of the
successive accumulation intervals N=1, 2, 3, and 4 are shown in
Fig.~\ref{lc3bands} by dashed vertical lines. Because of the small
number of counts in interval 3, spectra 3 and 4 were combined. The
time-integrated spectrum of the entire burst was accumulated over
the interval 0--256 ms. The raw count rate spectra were rebinned in
order to have at least 20 counts per energy bin and then fitted
using XSPEC, version 11.3 \citep{Arnaud1996}. The detector response
function was calculated specifically for the $55\fdg9$ burst
incidence angle, determined from the IPN localization data.

\begin{figure}
\centering
\includegraphics[]{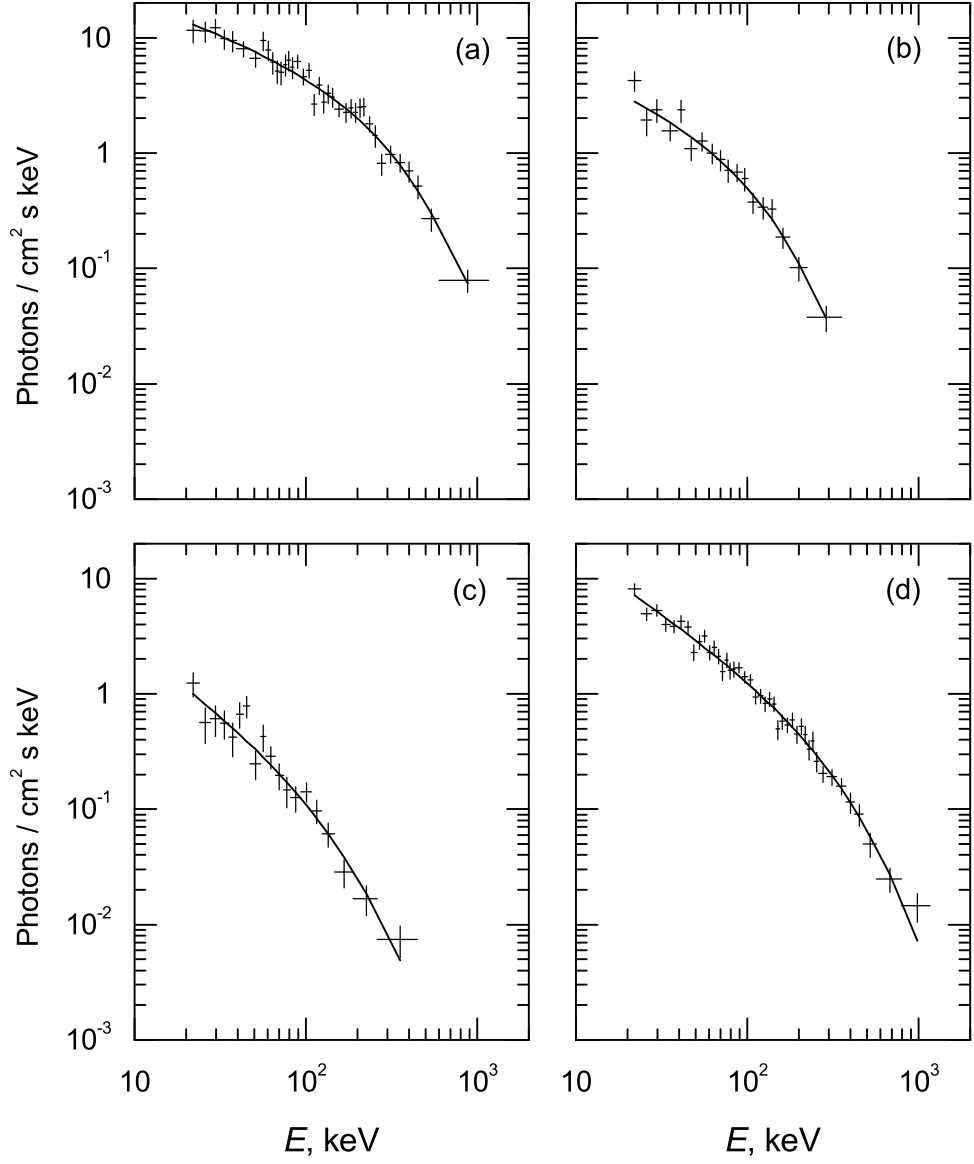}
\caption{Deconvolved photon spectra of the burst accumulated over
the (T-T$_0$) intervals 0-–64 ms (a), 64-–128 ms (b), 128-–256 ms
(c), and the time-integrated spectrum,  0-–256 ms (d).
\label{spectra}}
\end{figure}

A good fit was obtained for a power-law spectrum with an exponential
cutoff, $dN_{ph}/ dE \propto E^{-\alpha} \exp
\left[-(2-\alpha)E/E_p\right]$, where $E_p$ is the peak energy in
the $EF(E)$ spectrum. We also tested the Band (GRRM) model but no
statistically significant high energy power-law tail  was found in
any fitted spectrum. The deconvolved photon spectra for the
intervals 1, 2, and 3+4, and for the time-integrated spectrum of the
burst, are shown in Fig.~\ref{spectra}, and the spectral parameters
are summarized in Table~\ref{TableSpectra}. Comparison of the $E_p$
values shows that the spectrum for interval 1 is the hardest.
Returning to Fig.~\ref{lc3bands}, note that the accumulation time of
this spectrum is rather long (64~ms) compared to the characteristic
evolution time and hence the spectrum is strongly time-averaged.
Most of the high-energy photons are accumulated during the first
$\sim 20$~ms. This remarkable fact means that the flux of photons
with the hardest spectrum is emitted near the peak of the initial
pulse. The burst fluence in the 20~keV-–1.2~MeV range is
$2.00_{-0.26}^{+0.10} \times 10^{-5}$~erg~cm$^{-2}$. The 2-ms  peak
flux measured from T$_0$+0.016~s in the same energy band is
$1.61_{-0.50}^{+0.29} \times 10^{-3}$~erg~cm$^{-2}$~s$^{-1}$. All
the uncertainties are for the 90\% confidence level.

GRB~070201 was detected by Wind (Konus) with a time resolution of up
to 2~ms, by INTEGRAL (SPI-ACS) with a resolution of 50~ms, by
MESSENGER (MErcury Surface, Space ENvironment, GEochemistry, and
Ranging , Gamma-Ray and Neutron Spectrometer, \citet{Goldsten2007})
with a resolution of 1~s, and by the Swift BAT with 1~s resolution
(outside the coded field of view, J. Cummings, private communication
2007).  The initial triangulation was given in \citet{Hurley2007b}.
The coordinates of the center and vertices of the refined 446
arcmin$^2$ $3\sigma$ error box are listed in Table~\ref{TableBox}.
The center lies $\sim 1\degr$ away from the center of M31. Assuming
that the source of GRB~070201 is situated in M31 at a distance of
0.78~Mpc, the measured values of the fluence  and peak flux
correspond to an isotropic energy output  $Q = 1.5 \times
10^{45}$~erg, and an isotropic peak luminosity $L=1.2 \times
10^{47}$~erg~s$^{-1}$.

\begin{figure}
\centering
\includegraphics[]{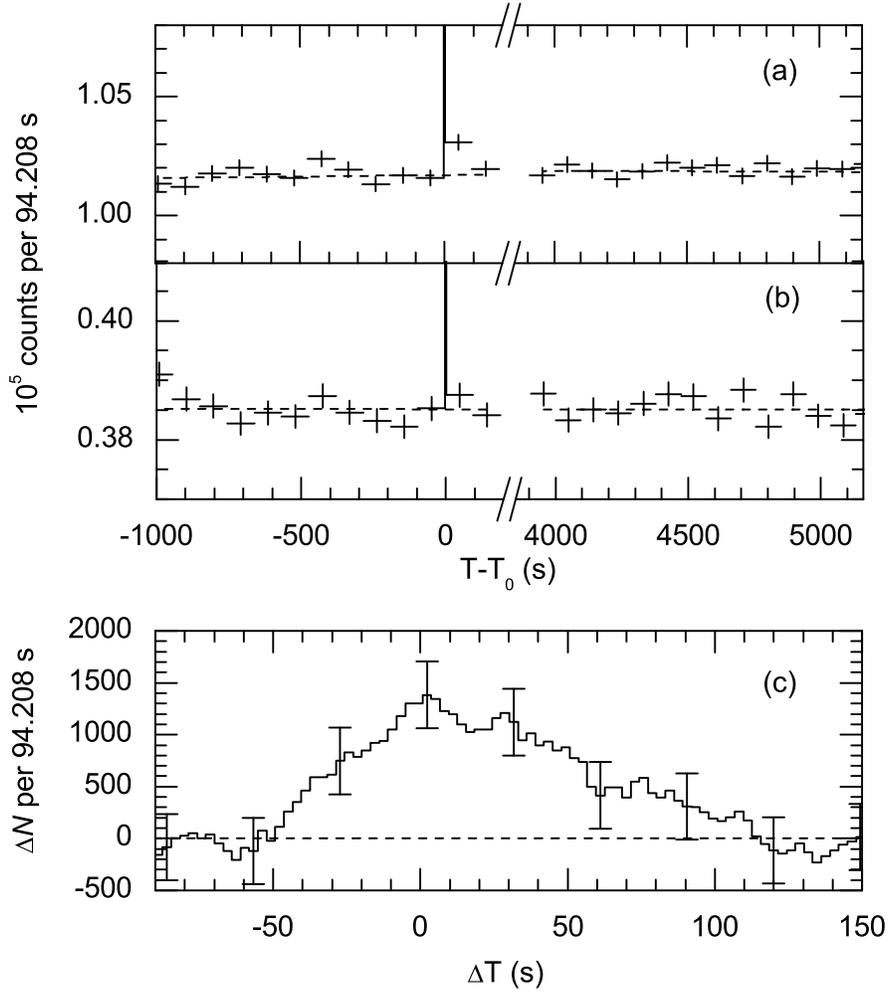}
\caption{Waiting mode count rates $N$ in the energy bands G1 (a) and
G2 (b) averaged over 94.208~s intervals (excluding the burst). For
G1, the first interval after T$_0$ exhibits a count excess $\Delta
N$ of 4.3$\sigma$ above the average background level. The bottom
panel (c) shows how this excess $\Delta N$ varies if the beginning
of the 94.208-interval is shifted by an interval $\Delta$T forwards
or backwards from T$_0$ step by step, with $\Delta$T=2.944~s.
\label{tail}}
\end{figure}

The relative proximity of M31 makes it worthwhile to search for the
afterglow of GRB~070201 in soft gamma rays, i.e., the tail of the
possible giant flare. While recording a burst in the trigger mode
with high time resolution, Konus also continues to measure the count
rates in each of the detectors S1 and S2 in four energy windows G1,
G2, G3, and Z  (a charged particles channel) in background mode with
a time resolution of 2.944~s. The initial pulse of GRB~070201 falls
completely in a single 2.944~s interval around T$_0$.  We began by
selecting a long series of count rate data, excluding this interval.
Then, in order to suppress fast statistical fluctuations we formed a
new series consisting of a sum over 94.2~s ($32 \times 2.944$s),
which can originate at the beginning of any 2.944~s interval. Two
such series, in the energy bands G1 and G2, are shown in
Figures~\ref{tail}a and b; they are synchronized to the first
2.944~s interval following the excluded one.  The position of the
excluded interval is marked  by a narrow vertical line. The
discontinuity at T-T$_0 \sim$240--3800~s  is due to the transfer of
the accumulated data to the on-board memory. A significant increase
in the count rate (4.3$\sigma$ above the average background level)
is seen only in the soft energy band G1 and only for the first large
interval after T$_0$. Panel~\ref{tail}c shows how the excess counts
$\Delta N$ vary if the beginning of the 94.2~s interval is shifted
forwards or backwards from T$_0$ step-by-step by 2.944~s.  The
dependence of $\Delta N$ on the time shift $\Delta$T implies that
the gradually decreasing soft gamma-ray flux is indeed present in
each 2.944~s interval after T$_0$ but is absent prior to it. This
important result confirms the detection of the decaying soft
afterglow of GRB~070201. For an OTTB spectrum with $kT \approx
30$~keV, the 94.2~s count excess $\Delta N = 1385 \pm 320$
corresponds to a fluence $S \approx 1 \times 10^{-6}$~erg~cm$^{-2}$.
Assuming that we are observing emission in the tail of a GF from
M31, its energy is $Q_{\mathrm{tail}} \approx 7 \times 10^{43}$~erg.

\section{DISCUSSION}
The Andromeda galaxy, M31, as the closest massive galaxy, has long
been regarded as one of the most likely candidates for searching for
observable  GFs from distant extragalactic SGRs \citep{Mazets1999b,
Bisnovatyi-Kogan1999}. It is well known that one of the dominant
features of the structure of M31 is a bright circular ring with a
radius of 10~kpc discovered by \citet{Arp1964} in H$\alpha$
observations. In agreement with numerous observations, the ring
reveals the main star-formation region of the galaxy. In the far
infrared, the first \textit{IRAS} images of M31 also revealed a
bright circular ring \citep{Habing1984}. The close correspondence
between the details of the H$\alpha$ and far infrared images of the
galaxy implies that the radiation of massive hot stars
\citep{Devereux1994} is the common energy source of these emissions.
Recently, high spatial resolution images of M31 have been obtained
in the mid- and far infrared at different wavelengths with the
\textit{Spitzer} space telescope \citep{Barmby2006, Gordon2006} and
in the far and near ultraviolet with \textit{GALEX
}\citep{Thilker2005}. These very similar images clearly show the
central region, the fragmented spiral arms, the main circular ring
with $R \approx 10$~kpc, and a less luminous outer ring with $R
\approx 14$~kpc.

\begin{figure}
\centering
\includegraphics[width=0.7\textwidth]{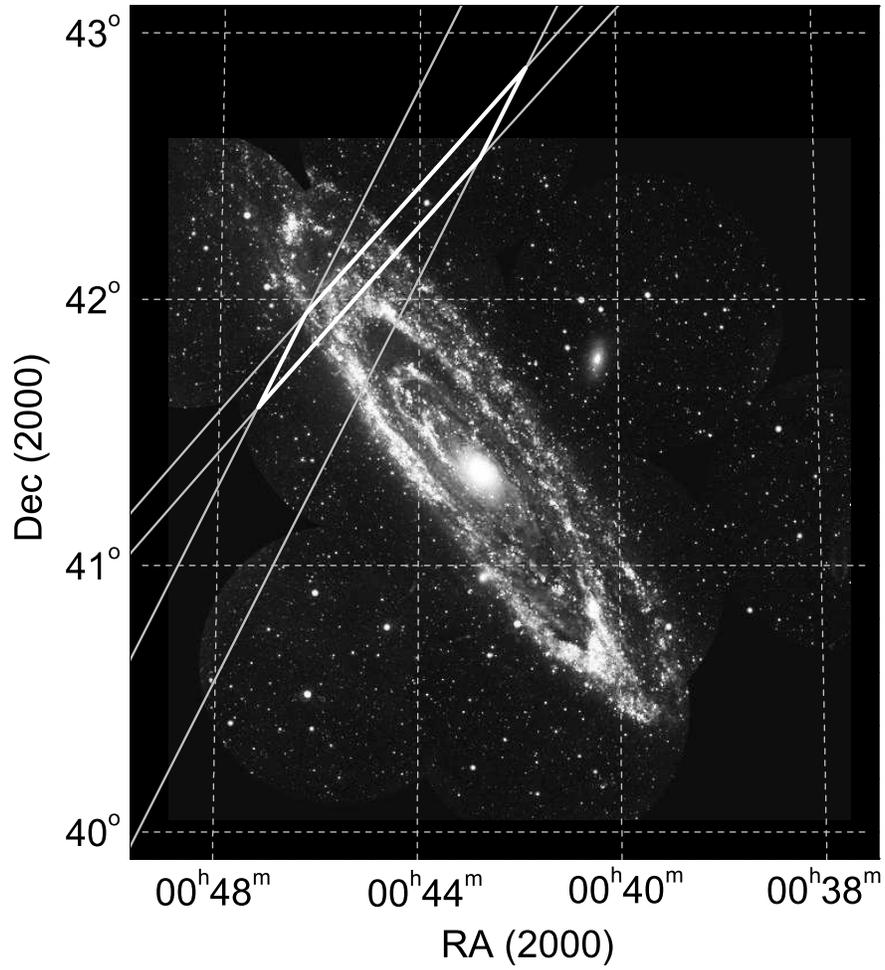}
\caption{UV image of the M31 galaxy
\citep{Thilker2005} and the 3$\sigma$ IPN error box of GRB~070201.
\label{localization}}
\end{figure}

Figure~\ref{localization} shows the UV image of M31
\citep{Thilker2005}. The IPN error box overlaps the northeastern
part of the galaxy. It contains segments of both the $R \approx
10$~kpc ring and of the weaker $R \approx 14$~kpc outer ring. The
probability of a chance overlapping of the error box and the image
is rather low ($\sim 10^{-4}$). Twenty-two sources from an
\textit{XMM-Newton} X-ray survey of M31 \citep{Pietsch2005} fall
within the error box. Among these, two sources are active galaxy
nuclei (AGN) and five are foreground objects. SGRs in the quiescent
state are typically sources of pulsating soft X-rays with an average
0.5--10~keV luminosity of $10^{35}$--$10^{36}$~erg~s$^{-1}$ and a
spectrum which can be described as a combination of a blackbody and
power law \citep{Kulkarni2003, Hurley2000, Kouveliotou2001}.
Spectral studies of the unidentified soft X-ray sources in the error
box which could be SGRs in M31 would clearly be important.

We now examine additional evidence which strengthens the
interpretation of the 2007 February~1 event as a GF from an SGR in
M31. This comes from a comparison of  the energetics, time history,
and spectral behavior of GRB~070201 with the corresponding
characteristics of the previously observed giant flares on 1979
March 5 (SGR~0525-66), 1998 June 18 (SGR~1627-41), 1998 August 27
(SGR~1900+14), and 2004 December 27 (SGR 1806-20).  Here we are
mainly interested in the characteristics of the GF initial pulses,
but unfortunately, it is just this information which is the hardest
to obtain reliably. When a sensitive gamma-ray detector records an
initial pulse with an immense intensity, it is strongly overloaded,
which makes it difficult or even impossible to obtain information
about the main part of the peak. If a giant pulse is recorded by
detectors of low-energy charged particles which have low sensitivity
to gamma-rays, there is no saturation problem in most cases.
However, severe difficulties are encountered instead when measuring
the photon energy, obtaining spectra, and consequently deriving
energy estimates. Nevertheless, the entire set of data on initial
pulses, obtained both with gamma-ray spectrometers
\citep{Mazets1979a, Mazets1982, Hurley1999a, Mazets1999a,
Frederiks2007a} and with charged-particle detectors
\citep{Hurley2005, Palmer2005, Terasawa2005, Schwartz2005,
Tanaka2007}, can be used to establish the general pattern of the
initial pulse.

We will restrict this comparison of GFs to a limited number of their
characteristics, for the following reasons. The time structure of
the initial pulse is likely to be very complex. The profile of both
the steep leading edge of the pulse, as in SGR~1806-20
\citep{Palmer2005}, and its more prolonged decay, as in SGR~1900+14
\citep{Tanaka2007}, contain significant intensity variations on both
short and long timescales. Among other things, the appearance of the
light curve will be strongly affected by the spectral variability of
the emission. At high count rates close to saturation level,
dead-time and pileup corrections of light curves and energy spectra
are difficult, if not impossible. Therefore, we include in
Table~\ref{TableSummary} only two fairly reliable characteristics of
the initial pulse, $t_R$ and $E_p$; $t_R$, the rise time, is the
time between the detection of hard gamma-radiation above the
background to the peak of the initial pulse, $E_p$ is the photon
energy at the maximum of the time-integrated $EF(E)$ spectrum of the
pulse. Table~\ref{TableSummary} also contains data on the peak
luminosity $L_{\mathrm{max}}$ and the energy release $Q$ of the
initial pulses, as well as the energy in the GF tails
$Q_{\mathrm{tail}}$. In keeping with tradition, we suggest the same
nomenclature for extragalactic and galactic SGRs, namely SGR
followed by the coordinates ($\alpha$, $\delta$) of the center of
the error box.

Table~\ref{TableSummary} shows that estimates of both the
luminosities and energy outputs in GFs are spread over a range of
several orders of magnitude.  On the other hand, the values of
$Q_{\mathrm{tail}}$ are practically equal.  Thus it is particularly
significant that the afterglow of GRB~070201 is similar to those of
the other SGRs. A small spread in SGR $Q_{\mathrm{tail}}$ values
implies similar magnetic field strengths, from confinement arguments
\citep{Thompson1995}. The only exception so far is the absence of a
tail in GF 980618 from SGR~1627-41, which is probably caused by the
relative weakness of the magnetic field or its orientation. On the
whole, Table 3 clearly demonstrates that both the temporal and
energetic characteristics of the event on 2007 February 1 match the
general pattern of a giant flare very closely. Beyond a doubt, we
can conclude that this event is a GF which originated  in
SGR~0044+42 in M31.

On the Russian side this work was supported by Federal Space Agency
of Russia and RFBR grant 06-02-16070. KH is grateful for IPN support
under NASA grants NNG06GE69G (the INTEGRAL guest investigator
program) and NNX07AR71G (MESSENGER Participating Scientist program).

\pagebreak

\clearpage
%

\clearpage

\clearpage

\clearpage

\clearpage

\clearpage
\begin{deluxetable}{rccc}
\tablewidth{0pt} \tablecaption{Summary of spectral fits
\label{TableSpectra}} \tablehead{ \colhead{Time interval} &
\colhead{$\alpha$} &
\colhead{$E_p$} & \colhead{$\chi^2$/dof}\\
\colhead{(s)} & \colhead{} & \colhead{(keV)} & \colhead{}}
\startdata
0--0.064 & $0.52_{-0.15}^{+0.13}$ & $360_{-38}^{+44}$ & 31.6/35\\
0.064--0.128 & $0.56_{-0.42}^{+0.38}$ & $128_{-16}^{+24}$ & 11.3/15\\
0.128--0.256 & $1.06_{-0.52}^{+0.42}$ & $123_{-25}^{+54}$ & 19.3/16\\
0--0.256 & $0.98_{-0.11}^{+0.10}$ & $296_{-32}^{+38}$ & 39.7/40\\
\enddata
\end{deluxetable}
\begin{deluxetable}{ccc}
\tablewidth{0pt} \tablecaption{IPN error box of GRB~070201
\label{TableBox}} \tablehead{ \colhead{} & \colhead{RA(2000)} &
\colhead{Dec(2000)}}
\startdata
Center & $00^{\mathrm h} 44^{\mathrm m} 32^{\mathrm s}$ &
$+42\arcdeg 14\arcmin 21\arcsec$\\
Vertices & & \\
1 & $00^{\mathrm h} 46^{\mathrm m} 18^{\mathrm s}$ & $+41\arcdeg
56\arcmin 42\arcsec$\\
2 & $00^{\mathrm h} 41^{\mathrm m} 51^{\mathrm s}$ & $+42\arcdeg
52\arcmin 08\arcsec$\\
3 & $00^{\mathrm h} 42^{\mathrm m} 47^{\mathrm s}$ & $+42\arcdeg
31\arcmin 41\arcsec$\\
4 & $00^{\mathrm h} 47^{\mathrm m} 14^{\mathrm s}$ & $+41\arcdeg
35\arcmin 35\arcsec$\\
\enddata
\end{deluxetable}

\begin{deluxetable}{lcccccc}
\tabletypesize{\scriptsize} \rotate \tablewidth{0pt}
\tablecaption{Some characteristics of known giant
flares\label{TableSummary}} \tablehead{\colhead{} & \colhead{SGR
0526-66\tablenotemark{a} } & \colhead{SGR 1627-41\tablenotemark{b}}
& \colhead{SGR 1900+14\tablenotemark{c}} & \colhead{SGR
1806-20\tablenotemark{d}} & \colhead{SGR 0952+69\tablenotemark{e}} &
\colhead{SGR 0044+42\tablenotemark{f}}\\
\colhead{} & \colhead{(LMC)} & \colhead{} & \colhead{} & \colhead{}
& \colhead{(M81)} & \colhead{(M31)}}
\startdata
Data, YYMMDD & 790305 & 980618 & 980827 & 041227 & 051103 &
070201\\
Distance, kpc & 50 & 10 & 15 & 15 & 3600 & 780\\
Rise time, $t_R$, ms & $\sim 25$ & $\sim 15$ & $\sim 15$ & $\sim 25$
& $\sim 6$ & $\sim 20$\\
Peak energy, $E_p$, keV & $\sim 500$ & $\sim 150$ & $> 250$ & $\sim
850$ & $\sim 900$ & $\sim 300$\\
Luminosity, $L_{\mathrm{max}}$, erg~s$^{-1}$ & $6.5 \times 10^{45}$
& $3.4 \times 10^{44}$ & $2.3 \times 10^{46}$ & $3.5 \times 10^{47}$
& $4.3
\times 10^{48}$ & $1.2 \times 10^{47}$\\
Energy release, $Q$, erg & $7 \times 10^{44}$ & $1 \times 10^{43}$ &
$4.3 \times 10^{44}$ & $2.3 \times 10^{46} $ & $7 \times 10^{46}$ &
$1.5
\times 10^{45}$\\
Tail energy, $Q_{\mathrm{tail}}$, erg & $3.6 \times 10^{44}$ &
absent & $1.2 \times 10^{44}$ & $2.1 \times 10^{44}$ & not detected
& $7 \times
10^{43}$\\
\enddata
\tablenotetext{a}{The previous dead-time correction for the event
\citep{Mazets1979a, Mazets1982} has been revised here, assuming that
the initial pulse displays the expected strong hard-to-soft spectral
evolution.  This resulted in a reduction in $t_R$ and an increase in
$L_{\mathrm{max}}$ and $Q$.}
\tablenotetext{b}{\citet{Mazets1999a, Aptekar2001}.}
\tablenotetext{c}{\citet{Tanaka2007, Hurley1999a, Mazets1999b}.}
\tablenotetext{d}{\citet{Palmer2005, Frederiks2007a}.}
\tablenotetext{e}{\citet{Frederiks2007b}.}
\tablenotetext{f}{This work.}
\end{deluxetable}

\begin{thebibliography}{}
\bibitem[Abbott et al.(2007)]{Abbott2007}Abbott, B., et al. 2007, ApJ,
submitted (astro-ph/0711.1163)

%
\bibitem[Aptekar et al.(2001)]{Aptekar2001}Aptekar, R. L., et al. 2001, \apjs, 137, 222
%
\bibitem[Arnaud(1996)]{Arnaud1996}Arnaud, K. A. 1996, in ASP Conf. Ser. 101,
Astronomical Data Analysis Software and Systems V, ed. G. Jacoby, \&
J. Barnes, (San Francisco: ASP), 17
%
\bibitem[Arp(1964)]{Arp1964}Arp, H., et al. 1964, \apj, 139, 1045
%
\bibitem[Atteia et al.(1987)]{Atteia1987}Atteia, J-L., et al. 1987, \apj, 320, L105
%
\bibitem[Barmby et al.(2006)]{Barmby2006}Barmby, P., et al. 2006, \apj, 650, L45
%
\bibitem[Bisnovatyi-Kogan(1999)]{Bisnovatyi-Kogan1999}Bisnovatyi-Kogan, G. S. 1999, preprint (astro-ph/9911275)
%
\bibitem[Cline et al.(1980)]{Cline1980}Cline,T., et. al. 1980, \apj, 237, L1
%
\bibitem[Devereux et al.(1994)]{Devereux1994}Devereux, N. A., et al. 1994, \aj, 108, 1667
%
\bibitem[Evans et al.(1980)]{Evans1980}Evans, W. D., et al. 1980, \apj, 237, L7
%
\bibitem[Feroci et al.(1999)]{Feroci1999}Feroci, M., et al. 1999, \apj, 515, L9
%
\bibitem[Frederiks et al.(2007a)]{Frederiks2007a}Frederiks, D. D., et al. 2007a, Astronomy Letters, 33, 1
%
\bibitem[Frederiks et al.(2007b)]{Frederiks2007b}Frederiks, D. D., et al. 2007b, Astronomy Lettrs, 33, 19
%
\bibitem[Goldsten et al.(2007)]{Goldsten2007}Goldsten, J., et al. 2007, \ssr, 131, 339
%
\bibitem[Golenetskii et al.(1984)]{Golenetskii1984}Golenetskii, S. V., et al. 1984, \nat, 307, 41
%
\bibitem[Golenetskii et al.(2005)]{Golenetskii2005}Golenetskii, S. V., et al. 2005, GCN Circular 4197
%
\bibitem[Golenetskii et al.(2007)]{Golenetskii2007}Golenetskii, S. V., et al. 2007, GCN Circular 6088
%
\bibitem[Gordon et al.(2006)]{Gordon2006}Gordon, K. D., et al. 2006, \apj, 638, L87
%
\bibitem[Habing et al.(1984)]{Habing1984}Habing, H. J., et al. 1984, \apj, 278, L59
%
\bibitem[Hurley et al.(1999a)]{Hurley1999a}Hurley, K., et al. 1999a, \nat, 397, 41
%
\bibitem[Hurley et al.(1999b)]{Hurley1999b}Hurley, K., et al. 1999b, \apj, 519, L143
%
\bibitem[Hurley et al.(2000)]{Hurley2000}Hurley, K., et al. 2000, in Gamma-Ray Bursts:
5th Huntsville Symp., ed. R. M. Kippen, R. S., Mallozi, \& G. F.
Fishman (Melville: AIP), 763
%
\bibitem[Hurley et al.(2005)]{Hurley2005}Hurley, K., et al. 2005, \nat, 434, 1098
%
\bibitem[Hurley et al.(2007a)]{Hurley2007a}Hurley, K., et al. 2007a, in preparation
%
\bibitem[Hurley et al.(2007b)]{Hurley2007b}Hurley, K., et al. 2007b, GCN Circular 6103
%
\bibitem[Kouveliotou et al.(1987)]{Kouveliotou1987}Kouveliotou, C., et al. 1987, \apj, 322, L21
%
\bibitem[Kouveliotou et al.(2001)]{Kouveliotou2001}Kouveliotou, C., et al. 2001, \apj, 558, L47
%
\bibitem[Kulkarni et al.(2003)]{Kulkarni2003}Kulkarni, S. R., et al. 2003, \apj, 585, 948
%
\bibitem[Laros et al.(1987)]{Laros1987}Laros, J. L., et al. 1987, \apj, 320, L111
%
\bibitem[Mazets et al.(1979a)]{Mazets1979a}Mazets, E. P., et al. 1979a, \nat, 282, 587
%
\bibitem[Mazets et al.(1979b)]{Mazets1979b}Mazets, E. P., et al. 1979b, Sov. Astron. Lett., 5(6), 343
%
\bibitem[Mazets et al.(1982)]{Mazets1982}Mazets, E. P., Golenetskii, S. V., Guryan,
Yu. A., \& Ilyinskii, V. N. 1982, \apss, 84, 173
%
\bibitem[Mazets et al.(1999a)]{Mazets1999a}Mazets, E. P., et al. 1999a, \apj, 519, L151
%
\bibitem[Mazets et al.(1999b)]{Mazets1999b}Mazets, E. P., et al. 1999b, Astronomy Letters, 25, 73
%
\bibitem[Palmer et al.(2005)]{Palmer2005}Palmer, D. M., et al. 2005, \nat, 434, 1107
%
\bibitem[Pietsch et al. (2005)]{Pietsch2005}Pietsch, W., Freyberg, M., \& Haberl, F. 2005, \aap, 434, 483
%
\bibitem[Schwartz et al.(2005)]{Schwartz2005}Schwartz, S. J., et al. 2005, \apj, 627, L129
%
\bibitem[Tanaka et al.(2007)]{Tanaka2007}Tanaka, Y. T., et al. 2007, \apj, 665, L55
%
\bibitem[Thompson \& Duncan(1995)]{Thompson1995}Thompson, C., \& Duncan, R. 1995, \mnras, 275,
255
%
\bibitem[Terasawa et al.(2005)]{Terasawa2005}Terasawa, T., et al. 2005, \nat, 434,
1110
%
\bibitem[Thilker et al.(2005)]{Thilker2005}Thilker, D. A., et al. 2005, \apj, 619, L67
%
\bibitem[Woods et al.(1999)]{Woods1999}Woods, P. M., et al. 1999, \apj, 519, L139
\end{thebibliography}
\end{document}